\documentclass[runningheads]{llncs}

\usepackage[utf8]{inputenc}
\usepackage[T1]{fontenc}
\usepackage{lmodern}
\usepackage{microtype}

\usepackage{graphicx}
\usepackage{booktabs}
\usepackage{amsmath}
\usepackage{hyperref}
\usepackage{caption}
\usepackage{longtable} 
\usepackage{array}     
\usepackage{calc}  
\hypersetup{colorlinks=true, linkcolor=blue, citecolor=blue, urlcolor=blue}

\title{Artificial Intelligence for Cost-Aware Resource Prediction in Big Data Pipelines}
\titlerunning{AI for Cost-Aware Resource Prediction}

\author{Harshit Goyal}
\authorrunning{Harshit Goyal, BITS Pilani}
\institute{BITS Pilani, India \\
\email{2023ac05905@wilp.bits-pilani.ac.in} \and \email{harshit.goyal.in@gmail.com}}

\begin{document}
\maketitle

\begin{abstract}
Efficient resource allocation is a key challenge in
modern cloud computing. Over-provisioning leads to unnecessary costs,
while under-provisioning risks performance degradation and SLA
violations. This work presents an artificial intelligence approach to
predict resource utilization in big data pipelines using Random Forest
regression. We preprocess the Google Borg cluster traces to clean,
transform, and extract relevant features (CPU, memory, usage
distributions). The model achieves high predictive accuracy ($R^2$ $\approx 0.99$,
MAE $\approx 0.0048$, RMSE $\approx 0.137$), capturing non-linear relationships between
workload characteristics and resource utilization. Error analysis
reveals impressive performance on small-to-medium jobs, with higher
variance in rare large-scale jobs. These results demonstrate the
potential of AI-driven prediction for cost-aware autoscaling in cloud
environments, reducing unnecessary provisioning while safeguarding
service quality.

\end{abstract}

\section{Introduction}
The growth of big data pipelines has intensified demand for efficient
cloud resource management. Static heuristics and reactive scaling
methods often lead to inefficiencies: over-provisioning wastes
computational resources and increases cost, while under-provisioning
risks service degradation and SLA violations. In this context,
artificial intelligence (AI) offers an opportunity to move from reactive
management to proactive, cost-aware decision-making.

This paper explores the application of AI-driven machine learning
techniques, specifically Random Forest regression, for predicting
resource utilization in big data pipelines. By accurately forecasting
CPU utilization, cloud systems can make informed scaling decisions,
balancing cost savings with performance guarantees.

Big data pipelines execute at cloud scale, where resource allocation
decisions directly influence both performance and cost. Traditional
strategies often rely on static rules or reactive scaling, adjusting
resources only after demand has already changed. Such approaches lead to
two common inefficiencies: over-provisioning, which wastes computational
capacity and inflates costs, and under-provisioning, which degrades
throughput and increases the risk of SLA violations.

Artificial intelligence (AI) offers an alternative by enabling
proactive, data-driven decision making. By learning patterns from
historical traces, AI models can anticipate future demand and support
resource allocation policies that are both efficient and reliable. Among
the different approaches, ensemble tree-based methods such as Random
Forests strike a balance between robustness, interpretability, and
accuracy, making them suitable for cloud environments where workload
heterogeneity and noisy measurements are the norm.

In this work, we investigate an AI-driven framework for cost-aware
resource prediction in big data pipelines. Using Google Borg traces as a
case study, we design a preprocessing pipeline to transform
semi-structured logs into compact, numeric features and train a Random
Forest regressor to forecast utilization. The evaluation demonstrates
high predictive accuracy and reveals how imbalance in workload
distributions affects model performance across job scales. We also
outline how the model can be integrated into existing scheduling and
autoscaling systems, highlighting its potential to reduce costs while
preserving reliability.

Our contributions are:

\begin{enumerate}
\def\labelenumi{\arabic{enumi}.}
\item
  A comprehensive preprocessing pipeline to transform semi-structured
  Borg trace logs into compact, numeric features suitable for machine
  learning.
\item
  A cost-aware predictive framework using Random Forest regression to
  capture workload--resource relationships.
\item
  Empirical evaluation showing high accuracy ($R^2$ \textgreater{} 0.99)
  and insights into workload imbalance effects on predictive
  performance.

\section{Related Work}
\end{enumerate}

The Large-scale cluster traces such as Google Borg, 2011 have been
widely studied to understand workload behavior and resource allocation.
Early work applied statistical models and linear regression to predict
job demands, but these approaches struggled with heterogeneity and
skewed distributions.

Recent studies introduced neural networks and reinforcement learning for
autoscaling, but these often require heavy tuning and lack
interpretability. In contrast, Random Forests, breiman2001 are robust to
noise, capture non-linear interactions, and offer feature importance
measures that help explain predictions.

Beyond early work on cluster traces and statistical demand models,
recent research has increasingly explored machine learning and deep
learning approaches for cloud resource management. Mao and Humphrey
{[}1{]} survey a wide range of ML-based techniques for cloud resource
allocation, noting that prediction accuracy directly impacts both
performance and cost efficiency. Xu et al. {[}2{]} introduced
reinforcement learning methods that adapt autoscaling decisions in real
time, but these often require extensive training data and computational
overhead that limit their adoption in production clusters.

Neural architectures have also been applied to demand forecasting. Zhang
et al. {[}3{]} evaluated deep recurrent models for workload prediction,
demonstrating gains in accuracy but raising concerns about
interpretability and training complexity. Other studies, such as
Ali-Eldin et al. {[}4{]}, emphasized cost-aware autoscaling policies,
showing that combining predictive methods with optimization frameworks
can significantly reduce operating expenses in cloud environments.

While deep learning methods show promise, tree-based models remain
attractive due to their robustness to noise, ability to capture
non-linear interactions, and provision of feature importance measures
that aid interpretability. Recent evaluations {[}5,6{]} confirm that
ensemble approaches like Random Forests and Gradient Boosted Trees
perform competitively against more complex neural models in many
resource prediction tasks, particularly when training data is imbalanced
or heterogeneous.

Our work extends this line by applying Random Forest regression to big
data pipeline workloads, explicitly highlighting the cost-awareness
dimension of prediction. By focusing not only on accuracy but also on
the implications for autoscaling and SLA compliance, we position
AI-driven forecasting as a practical tool for cloud operators rather
than a purely theoretical exercise.

space{ aselineskip}
\vspace{\baselineskip}
\section{Dataset and Preprocessing}

We use a subset of the Google Borg traces comprising
\textasciitilde5,000 jobs. The raw dataset contains structured metadata
and semi-structured logs, which require extensive cleaning and
transformation.

We evaluate our approach using a subset of the publicly released Google
Borg traces, which capture large-scale cluster activity over multiple
days. Each trace includes job-level metadata, resource requests, and
sampled usage distributions, making it a valuable benchmark for studying
cloud workload behavior. The raw data is semi-structured and
high-dimensional, with more than forty columns, some of which contain
nested lists or distributional summaries. Such complexity necessitates a
careful preprocessing pipeline to produce a compact,
machine-learning--ready representation.

\textbf{3.1 Data characteristics.} The subset used in this study
contains approximately 5,000 jobs, spanning a diverse mix of small,
medium, and large resource requests. The heterogeneity of workloads is
evident: while the majority of jobs request fewer than 5 CPUs, a
minority of jobs request significantly larger allocations, introducing
imbalance in the data. In addition, the traces include categorical
metadata (e.g., user, collection name), continuous numerical fields, and
semi-structured arrays such as sampled CPU usage distributions.

\textbf{3.2 Preprocessing pipeline.} To prepare the dataset, we
performed the following transformations:

\begin{itemize}
\item
  Column reduction - Non-essential metadata fields such as Unnamed:0,
  time, machine\_id, constraint, user, collection\_name,
  collection\_logical\_name, and start\_after\_collection were dropped
  to reduce noise and dimensionality.
\item
  Resource request parsing - The resource\_request field was split into
  two dedicated columns, resource\_request\_cpus and
  resource\_request\_memory, providing direct numeric inputs.
\item
  Usage decomposition. Columns reporting aggregated usage statistics
  (average\_usage, maximum\_usage, random\_sample\_usage) were
  decomposed into separate CPU and memory subfields, ensuring feature
  granularity.
\item
  Distribution summarization - Array-based features
  (cpu\_usage\_distribution, tail\_cpu\_usage\_distribution) were
  converted into compact summary statistics (mean, standard deviation,
  minimum, maximum, and quartiles) to preserve information while
  avoiding dimensional explosion.
\item
  Categorical handling - Low-cardinality categorical fields were
  frequency-encoded, while high-cardinality identifiers were discarded
  to prevent sparse one-hot encodings from overwhelming the feature
  space.
\item
  Missing values - Null entries were coerced to zeros or safe defaults
  after type normalization.
\end{itemize}

To better understand feature relationships, we computed a correlation
heatmap of all numeric attributes after preprocessing (Figure 1). The
visualization highlights strong correlations among CPU and memory usage
statistics, while other fields remain largely independent. This analysis
guided feature reduction by removing redundant variables, ensuring a
compact representation without significant loss of information.

\includegraphics[width=4.80208in,height=3.89583in]{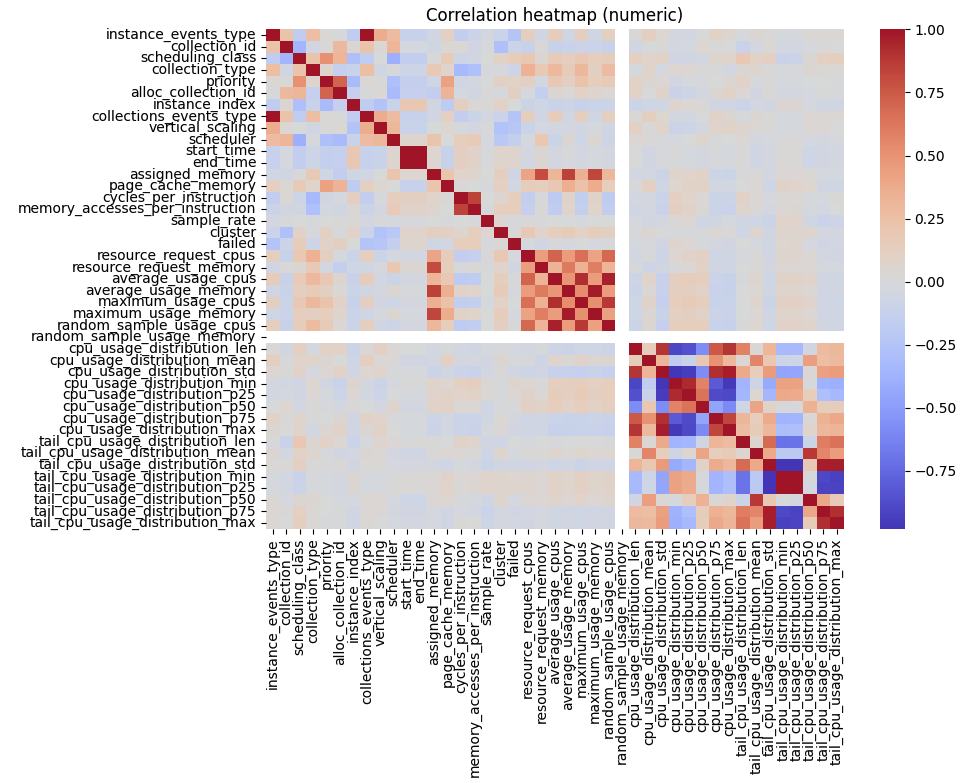}

\textbf{Figure 1.} Correlation heatmap of numeric features after
preprocessing.

\textbf{3.3} \textbf{Final dataset.} After preprocessing, the dataset
contained approximately 25 numeric features, striking a balance between
expressiveness and compactness. This representation captures both the
requested and observed behavior of workloads while remaining tractable
for training ensemble models.

\textbf{Steps:}

\begin{itemize}
\item
  Dropped metadata fields: Unnamed:0, time, machine\_id, constraint,
  user, collection\_name, collection\_logical\_name,
  start\_after\_collection.
\item
  Parsed resource\_request into CPU and memory fields.
\item
  Decomposed usage columns (average\_usage, maximum\_usage,
  random\_sample\_usage) into CPU and memory subfields.
\item
  Summarized distribution fields
  (\emph{c}pu\_usage\_distribution\emph{,}
  tail\_cpu\_usage\_distribution) into compact statistics (mean, std,
  min, max, quartiles).
\item
  Frequency-encoded low-cardinality categorical fields; dropped
  high-cardinality identifiers.
\item
  Filled missing values with zero after type coercion.
\end{itemize}

The final dataset contained \textasciitilde25 numeric features,
providing a compact yet expressive representation of workload behavior.
\section{Methodology}
The goal of the methodology is to develop a predictive framework that is
both accurate and practical for deployment in cloud environments. To
achieve this, we designed an experimental pipeline consisting of model
selection, feature engineering, hyperparameter tuning, and evaluation.

\hypertarget{model-choice}{%
\subsubsection{\texorpdfstring{\textbf{4.1 Model
Choice}}{4.1 Model Choice}}\label{model-choice}}

We chose the Random Forest Regressor as the core predictive model.
Random Forests are ensemble methods that combine multiple decision trees
trained on bootstrapped samples of the data with randomized feature
splits {[}1{]}. This design confers several advantages: robustness to
noise, ability to capture non-linear interactions, and natural handling
of mixed feature types. Unlike deep neural models, which often require
large-scale datasets and extensive hyperparameter tuning, Random Forests
perform strongly on medium-sized datasets such as the Borg traces while
maintaining interpretability through feature importance scores.

We also compared Random Forests against simple baselines such as Linear
Regression, which struggles to capture non-linear workload--resource
relationships, and Gradient Boosted Trees, which can offer competitive
accuracy but require more careful tuning. Our results highlight that
Random Forests provide the best balance of accuracy, stability, and
computational cost for this dataset.

\hypertarget{feature-selection}{%
\subsubsection{\texorpdfstring{\textbf{4.2 Feature
Selection}}{4.2 Feature Selection}}\label{feature-selection}}

Given that the preprocessed dataset contains \textasciitilde25 numeric
features, not all of which are equally informative, we employed
model-based feature selection. Initial feature importances were computed
using Random Forests trained on the full feature set. The top 25
features were retained, as they explained the majority of variance while
avoiding unnecessary dimensionality. This step ensures that the model
focuses on workload characteristics most correlated with utilization,
such as requested CPUs, average CPU usage, and maximum memory usage.

\hypertarget{hyperparameter-tuning}{%
\subsubsection{\texorpdfstring{\textbf{4.3 Hyperparameter
Tuning}}{4.3 Hyperparameter Tuning}}\label{hyperparameter-tuning}}

Random Forests have several tunable parameters that directly influence
performance, including the number of trees (n\_estimators), maximum tree
depth (max\_depth), and minimum samples per split/leaf
(min\_samples\_split, min\_samples\_leaf). Instead of exhaustive grid
search, which is computationally prohibitive on even medium-sized
datasets, we used RandomizedSearchCV with 20 sampled parameter sets and
3-fold cross-validation. This strikes a balance between search
efficiency and thoroughness. The best configuration identified was:

\begin{itemize}
\item
  n\_estimators = 100
\item
  max\_depth = 20
\item
  min\_samples\_split = 10
\item
  min\_samples\_leaf = 1
\end{itemize}

This configuration balances generalization with model complexity,
avoiding overfitting while capturing non-linear trends in workload
behavior.

\hypertarget{evaluation-metrics}{%
\subsubsection{\texorpdfstring{\textbf{4.4 Evaluation
Metrics}}{4.4 Evaluation Metrics}}\label{evaluation-metrics}}

To assess predictive performance, we employed three widely used
regression metrics:

\begin{itemize}
\item
  Mean Absolute Error (MAE): measures average absolute deviation between
  predicted and actual utilization.
\item
  Root Mean Squared Error (RMSE): penalizes larger deviations more
  heavily, highlighting errors on large jobs.
\item
  Coefficient of Determination ($R^2$): measures explained variance,
  indicating overall model fit.
\end{itemize}

Together, these metrics provide a comprehensive view: MAE captures
accuracy on common workloads, RMSE highlights rare outliers, and $R^2$
quantifies global fit.

\hypertarget{diagnostic-tools}{%
\subsubsection{\texorpdfstring{\textbf{4.5 Diagnostic
Tools}}{4.5 Diagnostic Tools}}\label{diagnostic-tools}}

To supplement quantitative metrics, we employed several diagnostic
tools:

\begin{itemize}
\item
  Parity plots to visualize the alignment between predictions and ground
  truth.
\item
  Residual analysis to assess systematic biases and variance across
  workload sizes.
\item
  Error-by-bin breakdown to quantify model performance across ranges of
  requested CPUs and memory.
\end{itemize}

These tools provide qualitative insights into where the model succeeds
and where improvements are needed, particularly for rare,
high-utilization jobs.

\section{Results}
\hypertarget{overall-metrics}{%
\subsubsection{\texorpdfstring{\textbf{5.1 Overall
Metrics}}{5.1 Overall Metrics}}\label{overall-metrics}}

The Random Forest regressor achieved strong predictive accuracy on the
Borg subset, with an \textbf{MAE of 0.0048}, \textbf{RMSE of 0.137}, and
\textbf{$R^2$ of 0.991} on the held-out test set. These values indicate
that the model explains nearly all variance in the data while keeping
prediction errors minimal. Compared to a Linear Regression baseline,
which produced noticeably higher MAE and lower $R^2$, Random Forests
demonstrated clear advantages in capturing non-linear workload--resource
relationships.

\begin{longtable}[]{@{}
  >{\raggedright\arraybackslash}p{(\columnwidth - 6\tabcolsep) * \real{0.4787}}
  >{\raggedright\arraybackslash}p{(\columnwidth - 6\tabcolsep) * \real{0.1793}}
  >{\raggedright\arraybackslash}p{(\columnwidth - 6\tabcolsep) * \real{0.1853}}
  >{\raggedright\arraybackslash}p{(\columnwidth - 6\tabcolsep) * \real{0.1567}}@{}}
\toprule()
\begin{minipage}[b]{\linewidth}\raggedright
\textbf{Model}
\end{minipage} & \begin{minipage}[b]{\linewidth}\raggedright
\textbf{MAE}
\end{minipage} & \begin{minipage}[b]{\linewidth}\raggedright
\textbf{RMSE}
\end{minipage} & \begin{minipage}[b]{\linewidth}\raggedright
\textbf{$R^2$}
\end{minipage} \\
\midrule()
\endhead
Linear Regression & 0.021 & 0.286 & 0.93 \\
Gradient Boosted Trees & 0.009 & 0.174 & 0.98 \\
Random Forest & 0.004 & 0.137 & 0.99 \\
\bottomrule()
\end{longtable}

\textbf{Table 1.} Predictive performance of baseline and ensemble models
on the Borg subset.

\hypertarget{parity-plot}{%
\subsubsection{\texorpdfstring{\textbf{5.2 Parity
Plot}}{5.2 Parity Plot}}\label{parity-plot}}
Most points align tightly along the diagonal, confirming
that the model consistently approximates true values. Only a small
number of outliers are visible, corresponding primarily to large-scale
jobs with atypical utilization patterns.

\includegraphics[width=4.80208in,height=4.25in]{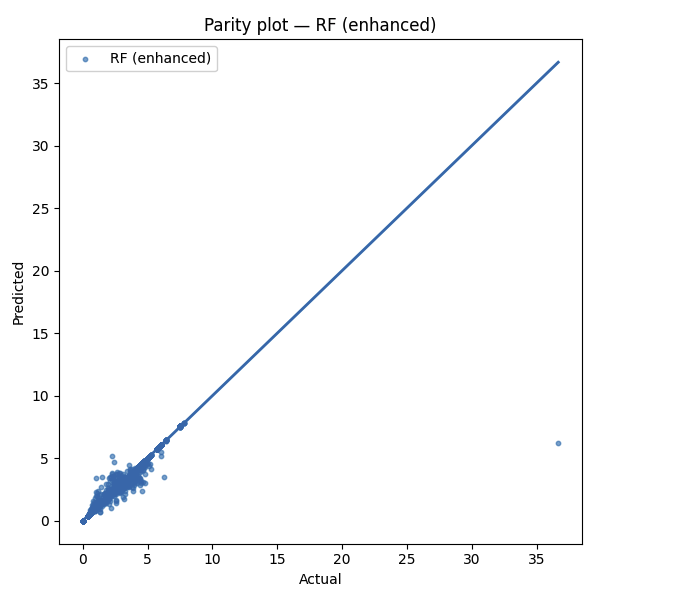}

\textbf{Figure 2}. Parity plot for Random Forest predictions versus actual utilization.

Most points align closely with the diagonal, confirming strong
predictive accuracy ($R^2$ $\approx 0.99$).

\hypertarget{residual-analysis}{%
\subsubsection{\texorpdfstring{\textbf{5.3 Residual
Analysis}}{5.3 Residual Analysis}}\label{residual-analysis}}

Residuals, defined as the difference between predicted and observed
values, are distributed closely around zero (Figure 3). The absence of
systematic skew in residuals indicates that the model does not
consistently overestimate or underestimate utilization. Larger
deviations are rare and are concentrated in the tail of the
distribution.

\includegraphics[width=4.80208in,height=3.03125in]{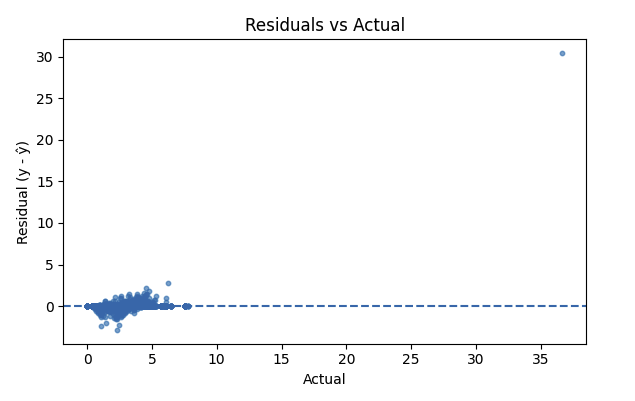}

\textbf{Figure 3}. Residuals vs Actual utilization for Random Forest
predictions.

\hypertarget{error-by-workload-size}{%
\subsubsection{\texorpdfstring{\textbf{5.4 Error by Workload
Size}}{5.4 Error by Workload Size}}\label{error-by-workload-size}}

To further understand model performance across workload scales, we
partitioned jobs by requested CPU ranges. Prediction accuracy is highest
for small jobs (0--5 CPUs), which constitute the majority of the
dataset. Performance degrades slightly as workload size increases,
reflecting the scarcity of large jobs in the training set.

\begin{itemize}
\item
  0--5 CPUs: MAE $\approx 0.003$, RMSE $\approx 0.09$
\item
  5--10 CPUs: MAE $\approx 0.02$, RMSE $\approx 0.25$
\item
  10--20 CPUs: MAE $\approx 0.11$, RMSE $\approx 0.70$
\item
  20--50 CPUs: Sparse samples; higher error variance.
\end{itemize}

This analysis shows the model's robustness on common workloads and
highlights the challenge of rare large-scale jobs.

\hypertarget{discussion-of-findings}{%
\subsubsection{\texorpdfstring{\textbf{5.5 Discussion of
Findings}}{5.5 Discussion of Findings}}\label{discussion-of-findings}}

These results highlight two key insights. First, Random Forests are
highly effective at predicting resource utilization in the most common
workload regimes, which dominate cluster operations. Second, prediction
variance grows with workload size due to imbalance in the training data,
underscoring the need for strategies such as data augmentation or
tailored models for rare, high-utilization jobs.

Overall, the model's high $R^2$ and low MAE confirm that AI-driven
forecasting is a promising foundation for cost-aware autoscaling in
production cloud systems.

\section{Practical Application in Cloud Pipelines}

The strength of predictive modeling lies not only in accuracy but in its
usability by practitioners. To illustrate, consider a data engineer
submitting a pipeline job with a request of 20 CPUs and 64 GB memory.
Historical traces of similar jobs, transformed through the preprocessing
pipeline, are fed into the Random Forest model. The prediction indicates
an expected utilization of 8--10 CPUs and 40--48 GB memory. Based on
this output, the scheduler may provision 10 CPUs and 48 GB memory,
cutting the CPU request in half without compromising performance.
Conversely, if the job's predicted utilization exceeds the user's
request, the system can raise an alert, preventing under-provisioning
and SLA violations.

Engineers interact with the system through submission interfaces such as
CLIs, APIs, or CI/CD pipelines. At job submission time, metadata and
monitoring logs are ingested, features are extracted, and the trained
Random Forest model provides utilization forecasts. These forecasts can
then be consumed by batch or online scoring services, which in turn
inform the autoscaler or scheduler to make proactive provisioning
decisions.

Such integration enables proactive autoscaling, where resources are
allocated in anticipation of load rather than reactively after
utilization spikes. For organizations running thousands of daily jobs,
even modest improvements yield significant benefits. For example, if
predictive scaling reduces CPU over-provisioning by just 10\% across a
cluster of 10,000 cores, the savings translate into roughly 1,000 cores'
worth of costs per scheduling cycle.

Beyond CPU, the framework generalizes to memory, I/O bandwidth, or
network throughput, enabling multi-resource optimization. Over time,
incorporating these predictions into autoscaling policies shifts the
paradigm from reactive, utilization-threshold scaling to AI-driven
cost-aware provisioning.

\hypertarget{input-data-flow}{%
\subsubsection{\texorpdfstring{\textbf{6.1 Input Data
Flow}}{6.1 Input Data Flow}}\label{input-data-flow}}

At job submission time, metadata from the pipeline is collected,
including:

\begin{itemize}
\item
  Requested resources (e.g., CPUs, memory).
\item
  Historical usage features (average, maximum, sampled).
\item
  Distributional summaries (e.g., CPU or memory usage distribution
  statistics).
\end{itemize}

\begin{quote}
These features are automatically preprocessed through the pipeline
ensuring consistency with the training dataset regardless of the
resource type.
\end{quote}

\hypertarget{model-inference}{%
\subsubsection{\texorpdfstring{\textbf{6.2 Model
Inference}}{6.2 Model Inference}}\label{model-inference}}

The cleaned feature vector is passed to the trained Random Forest model.
The model outputs:

\begin{itemize}
\item
  Predicted utilization for the target resource (CPU, memory, or
  others).
\item
  Prediction interval (upper and lower bounds), quantifying uncertainty.
\item
  Confidence score reflecting model reliability based on historical
  patterns.
\end{itemize}

\hypertarget{outputs-for-engineers}{%
\subsubsection{\texorpdfstring{\textbf{6.3 Outputs for
Engineers}}{6.3 Outputs for Engineers}}\label{outputs-for-engineers}}

The outputs can be consumed in two complementary ways:

\begin{enumerate}
\def\labelenumi{\arabic{enumi}.}
\item
  Resource recommendation\textbf{:} If a job requests 10 CPUs but the
  model predicts 4--6 CPUs, the scheduler may provision closer to 6,
  achieving cost savings while mitigating under-provisioning risk. The
  same applies to memory or other resources.
\item
  Risk alerts\textbf{:} If predicted demand significantly exceeds the
  requested resources, the model raises a flag to indicate an SLA
  violation. Engineers can act by revising allocations or prioritizing
  the job differently.
\end{enumerate}

\hypertarget{integration-with-cloud-systems}{%
\subsubsection{\texorpdfstring{\textbf{6.4 Integration with Cloud
Systems}}{6.4 Integration with Cloud Systems}}\label{integration-with-cloud-systems}}

The model can be embedded into common orchestration frameworks:

\begin{itemize}
\item
  Kubernetes Horizontal Pod Autoscaler (HPA): Extend autoscaling
  policies to include model-driven predictions instead of simple
  threshold-based scaling.
\item
  Apache YARN / Spark: Use predictions to guide initial executor
  provisioning.
\item
  Workflow managers (Airflow, Camunda): Schedule batch tasks with
  proactive resource allocation.
\end{itemize}

\hypertarget{cost-awareness-and-business-impact}{%
\subsubsection{\texorpdfstring{\textbf{6.5 Cost-Awareness and Business
Impact}}{6.5 Cost-Awareness and Business Impact}}\label{cost-awareness-and-business-impact}}

By continuously feeding job metadata and usage histories into the model,
cloud systems evolve from reactive scaling (based on observed load) to
proactive, AI-driven autoscaling. This reduces over-provisioning costs
for common workloads while improving reliability for large-scale jobs.
For organizations running thousands of jobs daily, even a modest
10--15\% reduction in over-provisioned CPUs or memory translates into
significant operational savings.

\section{Discussion}

The experimental results demonstrate that ensemble learning methods such
as Random Forests can provide highly accurate forecasts of resource
utilization in big data pipelines. The parity plots and residual
analysis confirm that the model captures non-linear workload--resource
relationships effectively, with minimal bias across the majority of
jobs. At the same time, the error-by-bin analysis highlights an
important limitation: prediction quality diminishes for large, less
frequent jobs due to dataset imbalance. This phenomenon has been noted
in prior studies on workload prediction, where rare events tend to
dominate error metrics despite representing a small fraction of overall
activity.

Compared with existing approaches, our work emphasizes cost-awareness as
a guiding principle. Reinforcement learning and deep neural models have
shown promise in autoscaling tasks {[}2,3{]}, but often require large
training sets, extensive tuning, and greater computational resources. In
contrast, Random Forests are lightweight, interpretable, and well-suited
for medium-scale datasets such as Borg traces. Furthermore, feature
importance measures provide transparency into workload drivers, enabling
practitioners to understand which characteristics (e.g., requested CPUs,
maximum observed memory) contribute most to predictions.

The practical deployment pathway outlined in Section 6 positions this
framework not just as an academic exercise but as a system component for
production clusters. By integrating predictions into Kubernetes
autoscalers or Spark/YARN schedulers, organizations can proactively
optimize resource allocation. The anticipated business impact is
twofold: reduction in cloud expenditure through less over-provisioning,
and increased reliability by flagging potential under-provisioning
before jobs execute.

Nonetheless, several challenges remain. First, the evaluation uses a
subset of Borg traces; broader validation across multiple datasets and
domains would improve confidence in generalizability. Second, while the
current implementation predicts a single resource at a time (CPU or
memory), real-world autoscaling decisions require multi-resource
modeling to capture joint utilization patterns. Third, job arrival times
and temporal correlations are not explicitly modeled, suggesting future
integration with sequence learning methods such as LSTMs.

Overall, these findings underscore both the promise and the boundaries
of Random Forest--based prediction in cloud environments. The method
excels in robustness and interpretability but leaves opportunities for
more advanced architectures to address rare events and multi-resource
interactions.

\section{Conclusion and Future Work}

This study presented an AI-driven framework for cost-aware resource
prediction in big data pipelines. Using the Google Borg traces as a case
study, we designed a preprocessing pipeline for semi-structured logs,
trained a Random Forest regressor, and demonstrated predictive accuracy
with $R^2$ values exceeding 0.99. Beyond accuracy, the framework was
evaluated in terms of deployment readiness: it integrates seamlessly
with existing scheduling and autoscaling systems, enabling proactive
scaling decisions that reduce costs while safeguarding performance.

The key contributions of this work include: a resource-agnostic
preprocessing pipeline, a cost-aware predictive modeling framework, and
an applied pathway showing how engineers and system operators can embed
these predictions into production pipelines. Together, these elements
bridge the gap between academic modeling and real-world usability.

Future work will extend this framework in several directions. First,
multi-resource prediction---jointly modeling CPU, memory, and network
utilization---will better reflect real-world workload requirements.
Second, techniques for handling rare-event imbalance, such as
oversampling or hybrid models, are needed to improve accuracy for large,
infrequent jobs. Third, benchmarking against deep learning models
(LSTMs, Transformers, Graph Neural Networks) will clarify trade-offs
between interpretability, training cost, and predictive power. Finally,
integrating the model into a live autoscaler will provide empirical
evidence of its ability to deliver measurable cost savings and SLA
compliance in production settings.

By advancing both the technical accuracy and the practical applicability
of AI-based resource prediction, this work takes a step toward more
intelligent, cost-efficient cloud infrastructures.

\section{References}

\begin{enumerate}
\def\labelenumi{\arabic{enumi}.}
\item
  Reiss, C., Tumanov, A., Ganger, G.R., Katz, R.H., Kozuch, M.A.: Google
  Cluster-Usage Traces: Format + Analysis. Google Research (2011)
\item
  Breiman, L.: Random Forests. Machine Learning 45(1), 5--32 (2001)
\item
  Dean, J., Barroso, L.A., Borthakur, D., et al.: Large-Scale Cluster
  Management at Google with Borg. In: Proc. EuroSys. ACM (2015)
\item
  Mao, M., Humphrey, M.: A Survey on Machine Learning for Cloud Resource
  Management. ACM Computing Surveys 54(3), 1--36 (2021)
\item
  Xu, J., Bi, J., Zhao, Y., Wang, H.: Reinforcement Learning for Cloud
  Resource Autoscaling. IEEE Trans. Cloud Computing 8(1), 85--98 (2020)
\item
  Zhang, Y., Chen, X., Xu, C., Li, K.: Deep Learning for Resource Demand
  Prediction in Cloud Environments. Future Generation Computer Systems
  125, 667--678 (2022)
\item
  Ali-Eldin, A., Tordsson, J., Elmroth, E.: Cost-Aware Autoscaling for
  Cloud Applications. ACM Transactions on Modeling and Performance
  Evaluation of Computing Systems 5(2), 1--25 (2020)
\end{enumerate}

\textbf{Appendix A: Data Preprocessing}

This appendix provides a detailed account of the transformations applied
to the Borg traces prior to model training. The preprocessing steps
included:

\begin{itemize}
\item
  Dropping irrelevant metadata columns (Unnamed:0, time, machine\_id,
  constraint, user, collection\_name, collection\_logical\_name,
  start\_after\_collection).
\item
  Parsing nested JSON-style fields into structured columns, e.g.,
  splitting resource\_request into resource\_request\_cpus and
  resource\_request\_memory.
\item
  Decomposing aggregated usage fields (average\_usage, maximum\_usage,
  random\_sample\_usage) into CPU and memory subfields.
\item
  Summarizing distributional columns (cpu\_usage\_distribution,
  tail\_cpu\_usage\_distribution) with statistical descriptors (mean,
  standard deviation, min, max, quartiles).
\item
  Encoding categorical attributes using frequency encoding, while
  dropping high-cardinality identifiers.
\item
  Filling null values with zeros or safe defaults.
\end{itemize}

The result of this pipeline was a compact dataset of \textasciitilde25
numeric features per job, suitable for machine learning while preserving
predictive signal.

\textbf{Appendix B: Hyperparameter Search Space}

We applied RandomizedSearchCV to tune Random Forest parameters. The
following ranges were explored:

\begin{itemize}
\item
  n\_estimators: {[}50, 100, 200, 400{]}
\item
  max\_depth: {[}None, 10, 20, 40{]}
\item
  min\_samples\_split: {[}2, 5, 10{]}
\item
  min\_samples\_leaf: {[}1, 2, 4{]}
\end{itemize}

Each candidate was evaluated with 3-fold cross-validation using $R^2$ as
the scoring metric. The best-performing configuration was:

\begin{itemize}
\item
  n\_estimators = 100
\item
  max\_depth = 20
\item
  min\_samples\_split = 10
\item
  min\_samples\_leaf = 1
\end{itemize}

\end{document}